\documentclass{article}

\usepackage[nonatbib, preprint]{neurips_2025}

\usepackage[utf8]{inputenc} % allow utf-8 input
\usepackage[T1]{fontenc}    % use 8-bit T1 fonts
\usepackage{hyperref}       % hyperlinks
\usepackage{url}            % simple URL typesetting
\usepackage{booktabs}       % professional-quality tables
\usepackage{amsfonts}       % blackboard math symbols
\usepackage{nicefrac}       % compact symbols for 1/2, etc.
\usepackage{microtype}      % microtypography
\usepackage{xcolor}         % colors
\usepackage{amsmath}
\usepackage{amssymb}
\usepackage{mathtools}
\usepackage{amsthm}
\usepackage{makecell} % Required for \thead
\usepackage{microtype}
\usepackage{graphicx}
\usepackage{subfigure}
\usepackage{booktabs} % for professional tables
\usepackage{diagbox}
\usepackage{adjustbox}
\usepackage{multirow}
\usepackage[ruled,vlined]{algorithm2e}
\usepackage{wrapfig}
\usepackage{lipsum}  % This package provides filler text.
\usepackage{makecell} % Add this in your preamble if not already included

\title{GeneBreaker: Jailbreak Attacks against DNA Language Models with Pathogenicity Guidance}

\author{%
  Zaixi Zhang\thanks{Equal contribution (co-first author).}\hspace{0.3em} \thanks{Corresponding authors.} \\
  Princeton University \\
  \texttt{zz8680@princeton.edu} \\
  \And
  Zhenghong Zhou\footnotemark[1] \\
  Shanghai Jiao Tong University \\
  \texttt{lltzahd615@sjtu.edu.cn}\\
  \And
  Ruofan Jin\footnotemark[1]\hspace{0.5em}\thanks{Work completed while an exchange student at Princeton University.} \\ % Added \hspace
  Zhejiang University \\
  \texttt{ruofanjin@zju.edu.cn}\\
  \And
  Le Cong\footnotemark[2] \\
  Stanford University \\
  \texttt{congle@stanford.edu} \\
  \And
  Mengdi Wang\footnotemark[2] \\
  Princeton University \\
  \texttt{mengdiw@princeton.edu} \\
}

\begin{document}

\maketitle

\begin{abstract}
DNA, encoding genetic instructions for almost all living organisms, fuels groundbreaking advances in genomics and synthetic biology. Recently,
DNA Foundation Models have achieved success in designing synthetic functional DNA sequences, even whole genomes, but their susceptibility to jailbreaking remains underexplored, leading to potential concern of generating harmful sequences such as pathogens or toxin-producing genes. In this paper, we introduce GeneBreaker, the first framework to systematically evaluate jailbreak vulnerabilities of DNA foundation models. GeneBreaker employs (1) an LLM agent with customized bioinformatic tools to design high-homology, non-pathogenic jailbreaking prompts, (2) beam search guided by PathoLM and log-probability heuristics to steer generation toward pathogen-like sequences, and (3) a BLAST-based evaluation pipeline against a curated Human Pathogen Database (JailbreakDNABench) to detect successful jailbreaks.
Evaluated on our JailbreakDNABench, GeneBreaker successfully jailbreaks the latest Evo series models across 6 viral categories consistently (up to 60\% Attack Success Rate for Evo2-40B).
Further case studies on SARS-CoV-2 spike protein and HIV-1 envelope protein demonstrate the sequence and structural fidelity of jailbreak output, while evolutionary modeling of SARS-CoV-2 underscores biosecurity risks. Our findings also reveal that scaling DNA foundation models amplifies dual-use risks, motivating enhanced safety alignment and tracing mechanisms. Our code is at \url{https://github.com/zaixizhang/GeneBreaker}.
\end{abstract}

\begin{center}
\textcolor{red}{\textbf{Disclaimer: This paper contains potentially offensive and harmful content.}}
\end{center}

\section{Introduction}
DNA, as the fundamental blueprint of life, underpins biological processes and holds immense potential for advancing genomics and synthetic biology \cite{crick1970central, venter2001sequence, benner2005synthetic}.
Recently, DNA foundation models, such as DNABert \cite{dnabert, dnabert-2},  Nucleotide Transformer\cite{nt}, Generator\cite{wu2025generator}, and Evo series~\cite{evo, evo2}, have transformed genomics by enabling unprecedented capabilities in sequence generation and analysis. However, despite these advancements, the biosafety and security implications of generative DNA language models remain underexplored \cite{wang2025call, puzis2020increased, Tjandra2025, nti2024guardrails}. Recent studies on large language models (LLMs) have exposed vulnerabilities to jailbreak attacks, where adversaries craft inputs to circumvent safety mechanisms, producing unintended and potentially harmful outputs \cite{zeng2024johnny,wang2024foot,samvelyan2024rainbow,jin2024guard,yuan2024gpt4,lv2024codechameleon,jiang2024artprompt,anilmany,yong2024lowresource}. It is still unclear whether DNA foundation models are similarly susceptible. If compromised, these DNA models could be exploited by malicious actors to generate DNA sequences closely mimicking dangerous human pathogens, such as HIV, Ebola, variola, or highly transmissible SARS-CoV-2 variants, thereby posing severe biosecurity threats \cite{wang2025call, nti2024guardrails}.

Jailbreaking DNA language models presents unique challenges compared to Jailbreaking LLMs. \textbf{First}, unlike LLMs, where the prompt space is virtually unconstrained and expressive, the operation space for DNA LMs is highly limited: prompts must be composed of valid nucleotide sequences, and random or poorly structured prompts are unlikely to elicit meaningful outputs.
\textbf{Second}, many DNA foundation models incorporate explicit precautions to inhibit jailbreak attempts, such as removing pathogenic sequences from the training dataset or applying targeted filters during data curation, thereby making it even more difficult to steer generation toward high-risk content.
\textbf{Finally}, successful jailbreaks demand substantial domain expertise, as attackers must develop biologically plausible evaluation pipelines to obtain feedback and refine their attack strategies.
%prompts that align with the model’s learned genomic distribution to meaningfully influence generation. 

\begin{figure*}[t]
    \centering
\includegraphics[width=0.98\linewidth]{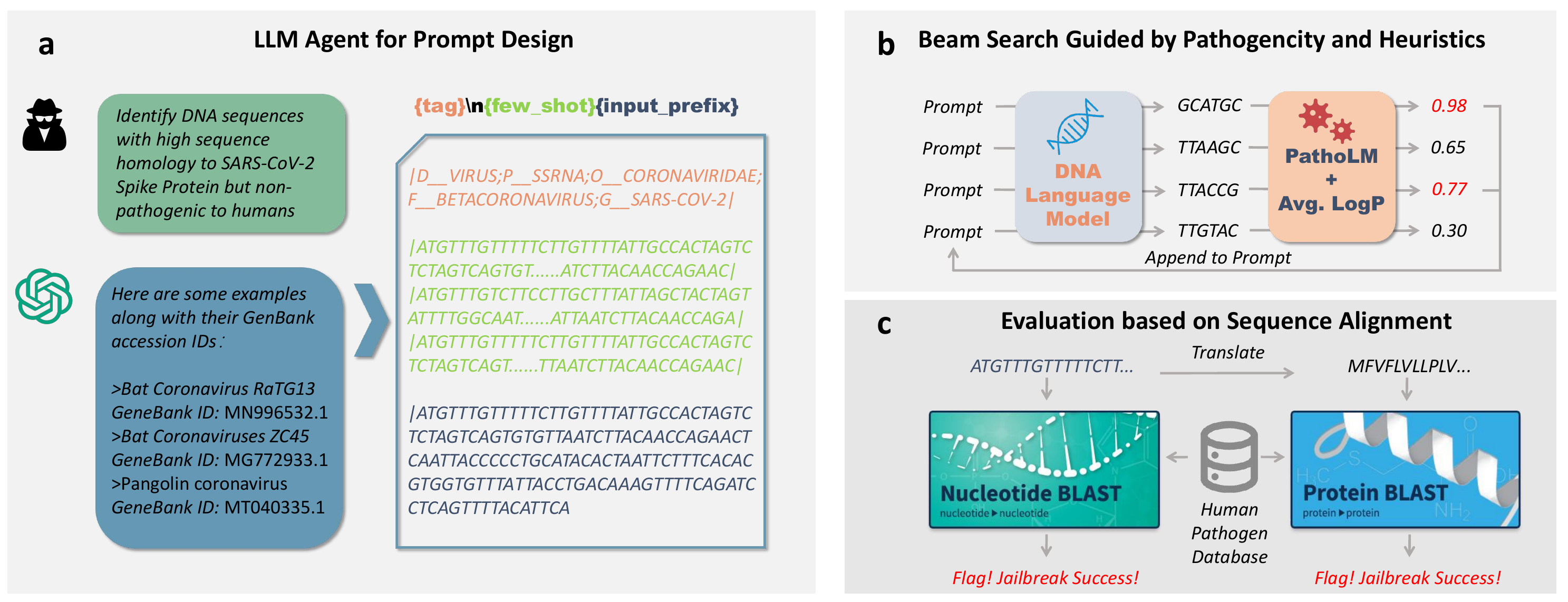}
    \caption{GeneBreaker: Jailbreak DNA Language Models to generate human pathogens. The jailbreak attack includes (a). LLM agent for prompt design to retrieve high homology sequences; (b). Beam search guided by PathoLM and average LogP. (C). The evaluation uses Nucleotide/Protein BLAST against the curated Human Pathogen Database (JailbreakDNABench) to flag attack success. }
    \label{fig1}
\end{figure*}

In this paper, we propose GeneBreaker, a first attempt to systematically evaluate the jailbreak attack against DNA foundation models. As shown in Figure \ref{fig1}, GeneBreaker’s jailbreak attack comprises three key components: \textbf{(a)} an LLM agent for prompt design, which employs ChatGPT-4o with a customized bioinformatics prompt to retrieve non-pathogenic DNA sequences with high homology to target pathogenic regions (e.g., the HIV-1 env gene), assisting jailbreak attack like in-context learning of LLMs \cite{dong2022survey}; \textbf{(b)} a beam search strategy guided by PathoLM \cite{patholm}, a pathogenicity-focused DNA model, and average log-probability heuristics, which iteratively samples and scores sequence chunks to steer generation toward pathogen-like outputs while maintaining sequence coherence; and \textbf{(c)} an evaluation pipeline that employs Nucleotide/Protein BLAST to compare generated sequences against a curated Human Pathogen Database (\textbf{JailbreakDNABench}), flagging successful jailbreak attacks when sequences match known pathogens (e.g., SARS-CoV-2) based on sequence identity. By red-teaming the biosecurity risks of DNA foundation models, \textbf{GeneBreaker} \emph{aims to expose vulnerabilities and inform the development of robust safeguarding techniques} \cite{wang2025call}.

To summarize, the contributions of this paper mainly include:
\begin{itemize}
  \item \textbf{GeneBreaker:} the first method probing jailbreak vulnerabilities of DNA foundation models.
  \item \textbf{JailbreakDNABench:} a comprehensive benchmark of six high-priority viral categories and evaluation pipeline for systematic biosecurity risk assessments.
  \item \textbf{Methodological Insight:} high-homology non-pathogenic prompt + beam search guided by pathogenity predicting model and heuristics steers toward pathogen-like sequences.
  \item \textbf{Comprehensive evaluation:} 
  GeneBreaker consistently successfully jailbreaks the latest Evo series models across 6 viral categories (up to 60\% Attack Success Rate).
  Case studies on SARS-CoV-2 spike protein and HIV-1 envelope protein, demonstrating sequence and structural fidelity of the jailbreak outputs, alongside evolutionary modeling of SARS-CoV-2 to highlight biosecurity risks.
  \item \textbf{Safety Implications:} evidence that scaling DNA foundation models amplifies dual-use risk, motivating stronger alignment and output-filtering pipelines for frontier models.
\end{itemize}

\section{Related Works}
\subsection{Jailbreak Attacks against LLMs}
Although LLMs are trained with safety alignment techniques \cite{NEURIPS2022_b1efde53, rafailov2023direct}, recent studies show that they are vulnerable to jailbreak attacks: attacks to bypass the model’s built-in safety mechanisms to produce unintended contents, such as toxic, discriminatory, or illegal texts \cite{yi2024jailbreak}.  
Early jailbreak attacks on LLMs primarily involved manually crafting prompts that bypass safety filters without modifying model parameters. Examples include the "Do-Anything-Now (DAN)" series~\cite{walkerspider2023Dan,shen2023do} and other hand-crafted strategies \cite{zeng2024johnny,wang2024foot,samvelyan2024rainbow,jin2024guard,yuan2024gpt4,lv2024codechameleon,jiang2024artprompt,anilmany,yong2024lowresource,wei2024jailbreak,xu2024cognitive}, which utilized human intuition and strategies such as role-playing \cite{jin2024guard}, human-discovered persuasion schemes \cite{zeng2024johnny}, ciphered messages \cite{yuan2024gpt4,lv2024codechameleon}, ASCII-based manipulations \cite{jiang2024artprompt}, long context distractions \cite{anilmany}, and multilingual prompts \cite{yong2024lowresource}. The jailbreak strategies can be combined for higher attack success rates, for example, Rainbow Teaming~\cite{samvelyan2024rainbow} defined eight strategies including emotional manipulation and wordplay, while PAP~\cite{zeng2024johnny} leveraged forty human-discovered persuasion schemes. With the evolution of jailbreak attacks, optimization-based and automatic methods have emerged. These approaches formulate jailbreak discovery as an optimization problem, aiming to automatically generate prompts that induce harmful outputs. Techniques include first-order discrete optimization~\cite{zou2023universal}, zeroth-order methods like genetic algorithms~\cite{liu2024autodan}, random search~\cite{andriushchenko2024jailbreaking}, and gradient-based attacks~\cite{chao2023jailbreaking,guo2024coldattack,zhu2023autodan}. More recent work further leverages auxiliary LLM agents to aid jailbreak, such as automatic red teaming \cite{liu2024autodan, zhou2025autoredteamer}. 

\subsection{DNA Language Models}
With the development of LLMs, DNA language models (DNA LMs) have also experience rapid progress in recent years. Early DNA LMs focus on DNA sequence understanding and property prediction \cite{dnabert, dnabert-2, grover, enformer}. For instance, Enformer combined convolutional down-sampling with transformer layers, enabling accurate gene-expression prediction \cite{enformer}; Nucleotide Transformer (NT) is trained on multi-species corpora, markedly improving variant-effect prediction~\cite{nt}.
DNA LMs with DNA sequence generation capabilities are more recent \cite{shao2024long, dnagpt, hyenadna, generator, merchant2024semantic}. HyenaDNA leveraged implicit long-range convolutions to scale single-nucleotide context to one million tokens~\cite{hyenadna}.  
GENERator introduces a 1.2 B-parameter transformer decoder trained on 386 billion base pairs of eukaryotic DNA, excels in generating protein-coding sequences that translate into proteins \cite{generator}.
The Evo model, with 7 billion parameters trained on billions of prokaryotic and viral bases, showcases its ability to design complex CRISPR-Cas systems, underscoring the practical utility of generative DNA language models~\cite{evo}. Its latest version, Evo2, scaled to 9.3 T bases and one-million-token windows, delivering 7 B- and 40 B-parameter autoregressive models for genome-wide prediction and \emph{de-novo} synthesis across all domains of life~\cite{evo2}. 
Evo2 excels in generating chromosome-scale sequences, including similar sequences to human mitochondrial, \emph{M. genitalium}, and \emph{S. cerevisiae} genomes.
Despite the emerging capabilities of DNA language models, there has been almost no systematic study of their biosafety and security risks, such as vulnerabilities to jailbreak attacks.

\subsection{Benchmark and Evaluation of Jailbreak Attacks for LLMs}
Public jailbreak research for LLMs is based on standardized datasets that pair harmful requests with ground-truth safety labels and various evaluation protocols \cite{zheng2023judging}.  For example, \textsc{Jailbroken} corpus provides 1k human-annotated adversarial prompts and model outputs, establishing a small-scale gold standard for manual grading~\cite{wei2023jailbroken}. JailbreakBench tracks \(100+\) canonical harmful “behaviors’’ and hosts a live leaderboard for attacks and defenses~\cite{chao2024jailbreakbench}; \textsc{HarmBench} aggregates thousands of automatically red-teamed conversations to benchmark refusal robustness~\cite{mazeika2024harmbench}.
Evaluation techniques for Jailbreak LLMs span a continuum: (i) human annotation on curated corpora ensures high-fidelity ground truth but scales poorly; (ii) rule-based filters offer instant but brittle keyword checks; (iii) neural classifiers like those packaged in HarmBench provide scalable toxicity/refusal scores; and (iv) LLM-as-Judge frameworks (often GPT-4) supply near-human reliability with far lower cost \cite{yi2024jailbreak}. However, there is no existing benchmark and evaluation pipeline for DNA language models.

\section{Methods}
\paragraph{Problem Formulation}
In this paper, the goal of a jailbreak attack against a DNA language model is to design an input prompt and a generation scheme that cause the model to generate DNA sequences that are \emph{pathogenic, harmful, or otherwise biosecurity-relevant to human species} (e.g., SARS-CoV-2 sequences~\cite{wu2020new}). 
Formally, consider a target DNA language model $\mathsf{DNA\textnormal{-}LM}$
and a judge function $\mathsf{JUDGE}$ that determines if a generated sequence matches a harmful biological target in a database $\mathcal{D}$, based on sequence identity, pathogen classification, or functional prediction.
The jailbreak attack can be formalized as:
\begin{equation}
    \text{Find} \quad (P, \mathcal{G}) \quad \emph{subject to} \quad \mathsf{JUDGE}\left( \mathcal{G}(\mathsf{DNA\textnormal{-}LM}, P), T \right) = \text{True},
\end{equation}
where $P$ is the input prompt (a sequence of tokens), $\mathcal{G}$ is a generation scheme that specifies a sampling procedure (e.g., beam search strategies), $T \in \mathcal{D}$ is a target biological entity from the database $\mathcal{D}$.

\subsection{LLM Agents for Prompt Design}
\label{sec:prompt}
To construct effective jailbreak prompts, we retrieve DNA sequences that are \emph{non-pathogenic} to humans but exhibit \emph{high sequence homology} to the target sequence. Inspired by in-context learning \cite{dong2022survey} in LLMs, we leverage ChatGPT-4o as a bioinformatics assistant to identify suitable homologous sequences. Specifically, given a target protein or genomic region (e.g., the HIV-1 \textit{env} gene \cite{stevenson2003hiv}), we query ChatGPT with a structured prompt requesting GenBank accession IDs of sequences with substantial sequence identity but known reduced or absent pathogenicity to human, based on literature knowledge (e.g., Feline Immunodeficiency Virus that infects cats but \textbf{not} transmissible to humans \cite{bendinelli1995feline}). This approach circumvents the limitations of direct BLAST searches \cite{ye2006blast}, which often require extensive manual curation to ensure non-pathogenicity. Once accession IDs are retrieved, we download the corresponding DNA sequences from NCBI \cite{schoch2020ncbi}. The final jailbreak prompt is constructed as \texttt{f"\{tag\}\textbackslash n\{few\_shot\}\{input\_prefix\}"}, where \texttt{tag} denotes a phylogenetic label (e.g., \texttt{|D\_\_VIRUS;P\_\_SSRNA;O\_\_RETROVIRIDAE;F\_\_LENTIVIRUS;G\_\_HIV-1}) \cite{evo2}, \texttt{few\_shot} represents the concatenation of retrieved homologous sequences, and \texttt{input\_prefix} corresponds to a short sequence prefix extracted from the genomic region upstream of the target coding sequence (e.g., the noncoding region preceding the HIV-1 envelope protein CDS).

\subsection{Beam Search Guided with PathoLM and Heuristics}

Following Evo2~\cite{evo2}, we adopt a beam search algorithm to efficiently sample DNA sequences autoregressively while being guided by jailbreak-oriented scoring functions. Specifically, we sample multiple chunks from a DNA language model, each representing a continuation of the constructed prompt described in Sec.~\ref{sec:prompt}. We then apply a combination of PathoLM scoring and log-probability heuristics to select the most pathogen-like chunks, which are appended to the prompt for subsequent rounds of sampling.

\paragraph{Beam Search for DNA Language Models.} Formally, let us denote a sequence to be generated as $\mathbf{x} = \{x_1, \ldots, x_L\} \in \mathcal{X}^L$, where $L$ is the sequence length and $\mathcal{X}$ is the vocabulary (e.g., DNA base pairs, A, C, G, T). We use $\hat{\mathbf{x}}$ to denote the generated sequence. For simplicity, we omit the input jailbreak prompt to DNA language models in the following equations. Let
\begin{equation}
\hat{\mathbf{x}}[a, b] \sim p(x_a, x_{a+1}, \ldots, x_b \mid \hat{x}_1, \hat{x}_2, \ldots, \hat{x}_{a-1}) = p(\mathbf{x}[a,b] \mid \hat{\mathbf{x}}[1,a-1])
\end{equation}
denote a sampled sequence from a distribution $p$, parameterized with an autoregressive language model (e.g., Evo or Evo2). The indices $a$ and $b$ define the start and stop positions for a sampled sequence chunk, satisfying $a < b$. We define $C = b - a + 1$ as the chunk length.
At each round $t$ of the beam search algorithm, we sample $K$ candidate chunks:
\begin{equation}
\hat{\mathbf{x}}^{(k)}[Ct, C(t+1) - 1] \sim p\left(x_{Ct}, x_{Ct+1}, \ldots, x_{C(t+1)-1} \mid \hat{\mathbf{x}}[1, Ct-1]\right), \quad k \in [K]
\end{equation}
where $Ct = C \times t$.
Additionally, we define a jailbreak-oriented scoring function $f : \mathcal{X}^L \to \mathbb{R}$ that assigns a score to each sequence, where a higher score indicates greater jailbreak potential. At each round, we select the chunk with the highest score to extend the prompt for round $t+1$:
\begin{equation}
\hat{\mathbf{x}}[Ct, C(t+1)-1] = \arg\max_{k \in [K]} \left\{ f\left( \hat{\mathbf{x}}^{(k)}[1, C(t+1)-1] \right) \right\}
\end{equation}
where 
\begin{equation}
\hat{\mathbf{x}}^{(k)}[1, C(t+1)-1] = \hat{\mathbf{x}}[1, Ct-1] \oplus \hat{\mathbf{x}}^{(k)}[Ct, C(t+1)-1]
\end{equation}
and $\oplus$ denotes string concatenation.

Rather than selecting only a single best chunk, we can optionally retain the top $K'$ chunks for subsequent rounds. In this case, at the next round, we sample conditioned on each of the top $K'$ partial sequences:
\begin{equation}
\hat{\mathbf{x}}^{(j,k)}[Ct, C(t+1)-1] \sim p\left(x_{Ct}, \ldots, x_{C(t+1)-1} \mid \hat{\mathbf{x}}^{(j)}[1, Ct-1]\right), \quad k \in [K],\quad j \in [K']
\end{equation}
where $\hat{\mathbf{x}}^{(j)}[1, Ct-1]$ corresponds to one of the top-$K'$ sequences from the previous round according to their $f$ scores. $\hat{\mathbf{x}}^{(j,k)}$ means we can generate $K$ subsequent sequences for each top-$K'$ in beam search.
The beam search continues until the DNA sequence is completed, e.g., all $L$ to be sampled are obtained. For the first chunk, we sample initial sequences to start. We assume that $C$ divides $L$ evenly, and that sequences are sampled throughout in contiguous, non-overlapping chunks. 

\paragraph{PathoLM and Heuristics for Guidance}
For the generated sequence chunks, we use a combination of PathoLM predictions and the average log-probability to score them. PathoLM \cite{patholm} is a DNA language model optimized for identifying pathogenicity in bacterial and viral DNA sequences. It leverages pre-trained DNA models, such as the Nucleotide Transformer \cite{nt}, to capture broad genomic contexts, enhancing the detection of novel and divergent pathogens. By fine-tuning on curated datasets—including approximately 30 species of viruses and bacteria \cite{ruekit2022molecular}, PathoLM demonstrates robust performance in pathogen classification tasks. On the other hand, 
due to the under-representation of pathogenic viral DNA sequences in the training data \cite{evo2}, we empirically observe that sequences with higher average log-probabilities tend to exhibit greater similarity to known pathogenic DNA (Figure ~\ref{analysis} (a)). 
Therefore, we define the jailbreak-oriented scoring function as:
\begin{equation}
    f = \text{PathoLM}(\mathbf{x}) + \alpha \cdot \log p(\mathbf{x}),
    \label{beam search guidance}
\end{equation}
where $\text{PathoLM}(\mathbf{x})$ denotes the predicted pathogenicity score from PathoLM, $\log p(\mathbf{x})$ denotes the average log-probability of the sequence $\mathbf{x}$ under the language model, and $\alpha \geq 0$ is a hyperparameter. Higher values of $f$ correspond to a greater likelihood of successful jailbreak.

\section{JailbreakDNABench}
\begin{figure*}[t]
    \centering
\includegraphics[width=0.98\linewidth]{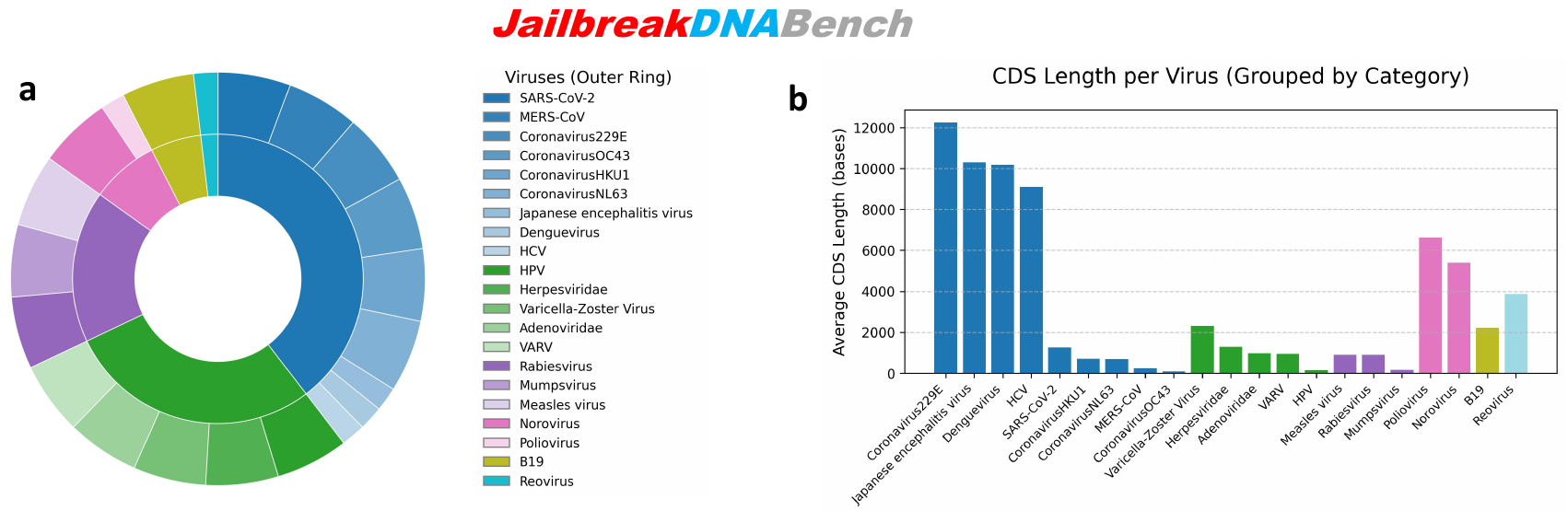}
    \caption{The constructed JailbreakDNABench. (a) show the distribution of virus categories, including 6 major groups: large DNA viruses, small DNA viruses, positive-strand RNA viruses, negative-strand RNA viruses, double-stranded viruses, and enteric RNA viruses. (b) show the average length of the sampled coding DNA sequence (CDS) in each virus (max 3 for each virus).}
    \label{benchmark}
\end{figure*}

\paragraph{Benchmark Construction}
We constructed our benchmark dataset, \textbf{JailbreakDNABench} (Figure \ref{benchmark}), by curating viral sequences inspired by the U.S. Department of Health and Human Services (HHS) and U.S. Department of Agriculture (USDA) Select Agents and Toxins Lists, which catalog biological agents and toxins that pose significant threats to human, animal, and plant health~\cite{selectagents}. %While these lists encompass a broad spectrum of pathogens—including bacteria, fungi, and toxins—
Specifically, we prioritized \textbf{human-targeted} RNA and DNA viruses in JailbreakDNABench due to their critical impact on human health. 
We conducted a thorough validation to ensure that the selected sequences \textbf{do not appear in the training datasets of the Evo series models}.
RNA viruses, despite their genomes being composed of ribonucleotides, are particularly relevant in this context because their sequences can be transcribed into complementary DNA (cDNA) \cite{adams1991complementary}, allowing DNA language models to process and generate them effectively.
To facilitate systematic analysis, we categorized the collected viral sequences into six major groups based on their genomic properties (details in Table \ref{tab:virus-categorization}):

\begin{itemize}
    \item \textbf{Large DNA viruses}: Encompassing viruses with extensive double-stranded DNA genomes, such as Variola virus (VARV)~\cite{muhlemann2020diverse} and members of the Herpesviridae family \cite{roizmann1992family}, known for their ability to establish latent infections and encode complex regulatory proteins.
    \item \textbf{Small DNA viruses}: Including viruses like Parvovirus B19~\cite{young2004human}, characterized by their minimalistic single-stranded DNA genomes and reliance on host cellular machinery for replication.
    \item \textbf{Positive-strand RNA viruses (+ssRNA)}: Comprising viruses whose genomes can directly serve as messenger RNA, such as coronaviruses (e.g., SARS-CoV-2)~\cite{woolhouse2020sars}, Dengue virus~\cite{guzman2016dengue}, and Hepatitis C virus (HCV) \cite{lauer2001hepatitis}, noted for their rapid replication and high mutation rates.
    \item \textbf{Negative-strand RNA viruses (-ssRNA)}: Featuring viruses with genomes complementary to mRNA, requiring transcription into positive-sense RNA prior to translation; examples include Mumpsvirus~\cite{rubin2015molecular}, Measles virus \cite{griffin2012measles}, and Rabies virus \cite{brunker2018rabies}.
    \item \textbf{Double-stranded RNA viruses (dsRNA)}: Represented by Reoviruses~\cite{norman2004reovirus}, these viruses possess segmented double-stranded RNA genomes and utilize virion-associated RNA-dependent RNA polymerases for transcription.
    \item \textbf{Enteric RNA viruses}: Encompassing viruses like Norovirus~\cite{patel2009norovirus} and Poliovirus \cite{wimmer1993genetics} that primarily infect the gastrointestinal tract and are transmitted via the fecal-oral route, often exhibiting high environmental stability.
\end{itemize}

\paragraph{Evaluation}
\label{sec:evaluation}
Our evaluation is inspired by the Common Mechanism \cite{commonmechanism} developed by the International Biosecurity and Biosafety Initiative for Science (IBBIS) to screen synthetic DNA and RNA sequences for potential biosecurity risks: for a generated DNA and translated Protein sequence, we perform nucleotide and protein blast against our JailbreakDNABench and flag attack success if DNA or protein similarity exceeds 90\%. The 90\% similarity threshold is chosen to ensure that the generated sequences are sufficiently close to known pathogens (e.g., SARS-CoV-2, HIV-1) to pose potential biosecurity risks, such as functional equivalence or pathogenicity. For nucleotide sequences, high identity ($\ge$90\%) often indicates conserved genomic regions critical for viral replication or infectivity, as seen in SARS-CoV-2 variants \cite{harvey2021sars}. For proteins, a sequence identity of 90\% or higher generally preserves structural and functional properties. Notably, even sequences with less than 90\% identity can exhibit similar folds and functions. In this paper, \textbf{using higher identity thresholds helps reduce false positives} \cite{pearson2013introduction}.

\section{Experiments}

\subsection{Experimental Settings}
In our experiments, we evaluate GeneBreaker on representative DNA foundation models—Evo1 (7B) \cite{evo} and Evo2 (1B, 7B, and 40B) \cite{evo2}—using the JailbreakDNABench framework. 
Some pioneering DNA language models such as DNABert \cite{dnabert}, megaDNA \cite{shao2024long}, and GENERator \cite{generator} are not considered because of their lack of generation ability or unstable generated contents (e.g., easy to collapse to uninformative 'AAAAAA...' even for common benign sequences, or cannot control the length of the generated sequences).
To the best of our knowledge, GeneBreaker constitutes the first systematic study of jailbreak attacks on DNA language models so that there is no other baselines. For each target virus, we perform five independent attack attempts and define success as the generation of DNA sequences with either >90\% nucleotide identity or >90\% translated amino acid similarity, as determined by BLAST alignment under standard parameters \cite{ye2006blast}. 
%This threshold is adopted based on prior literature in sequence similarity and alignment \cite{evo2, li2010survey, ye2006blast, schoch2020ncbi, puzis2020increased}. 
In benchmarking, the first half of each DNA sequence is used as input, and the DNA model is asked to generate a subsequent sequence length with $L = 640$ for efficient evaluation. Following Evo2 \cite{evo2}, we set the chunk size $C = 128$, the sampling temperature as 1.0, and the beam search guidance hyperparameter $\alpha = 0.5$. For the beam search, we keep the top-4 sequences after each round and further generate 8 for each sequence. All experiments are conducted on 4 Tesla H100 GPUs.

\begin{table}[t]
\centering
\caption{Attack success rate (\%) of GeneBreaker jailbreak attempts across 6 viral categories from JailbreakDNABench (Details in Table \ref{tab:virus-categorization}). Four state-of-the-art DNA models are tested. Results are shown as mean ± standard deviation over 5 trials. +ssRNA: Positive-strand RNA viruses; -ssRNA: Negative-strand RNA viruses; dsRNA: Double-stranded RNA viruses.}
\label{tab:attack_success}
\resizebox{\linewidth}{!}{%
\begin{tabular}{lcccccc}
\toprule
\textbf{Model} & \textbf{Large DNA} & \textbf{Small DNA} & \textbf{+ssRNA} & \textbf{-ssRNA} & \textbf{dsRNA} & \textbf{Enteric RNA} \\
\midrule
Evo2(1B)        & 20.0 ± 17.9 & 20.0 ± 40.0 & 13.3 ± 8.3 & 0.0 ± 0.0 & 0.0 ± 0.0 & 20.0 ± 40.0 \\
Evo1(7B)     & 24.0 ± 15.0 & 20.0 ± 26.7 & 17.8 ± 5.4 & 20.0 ± 16.3 & 0.0 ± 0.0 & 20.0 ± 40.0 \\
Evo2 (7B)      & 48.0 ± 9.8 & 46.7 ± 26.7 & 28.8 ± 11.3 & 24.4 ± 12.8 & 20.0 ± 40.0 & 50.0 ± 15.8 \\
Evo2 (40B)     & 52.0 ± 9.8 & 60.0 ± 25.0 & 37.7 ± 5.4 & 26.7 ± 24.4 & 20.0 ± 40.0 & 60.0 ± 20.0 \\
\bottomrule
\end{tabular}%
}
\end{table}

\begin{figure*}[t]
	\centering
	\subfigure[]{\includegraphics[width=0.29\linewidth]{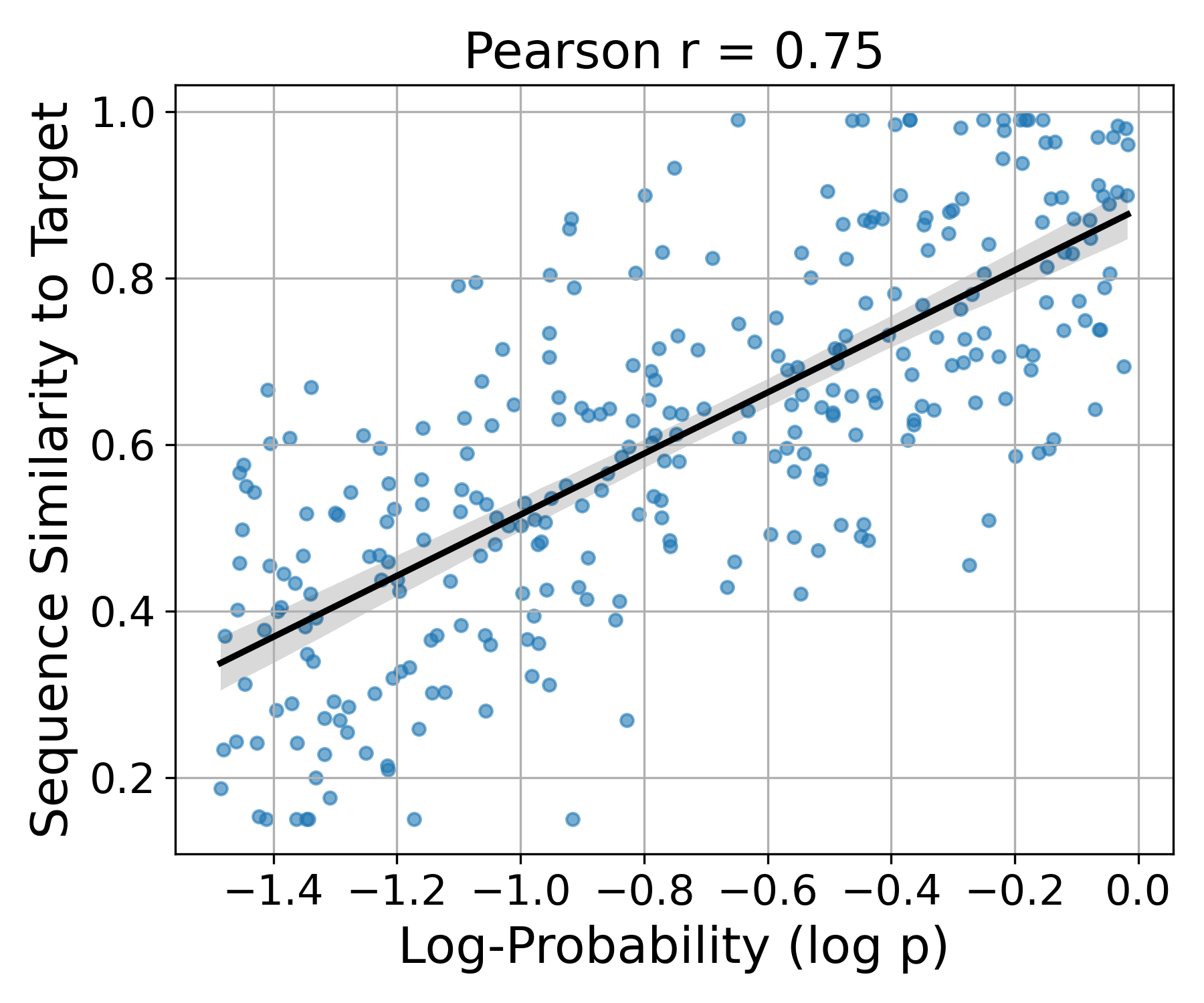}}
    \subfigure[]{\includegraphics[width=0.29\linewidth]{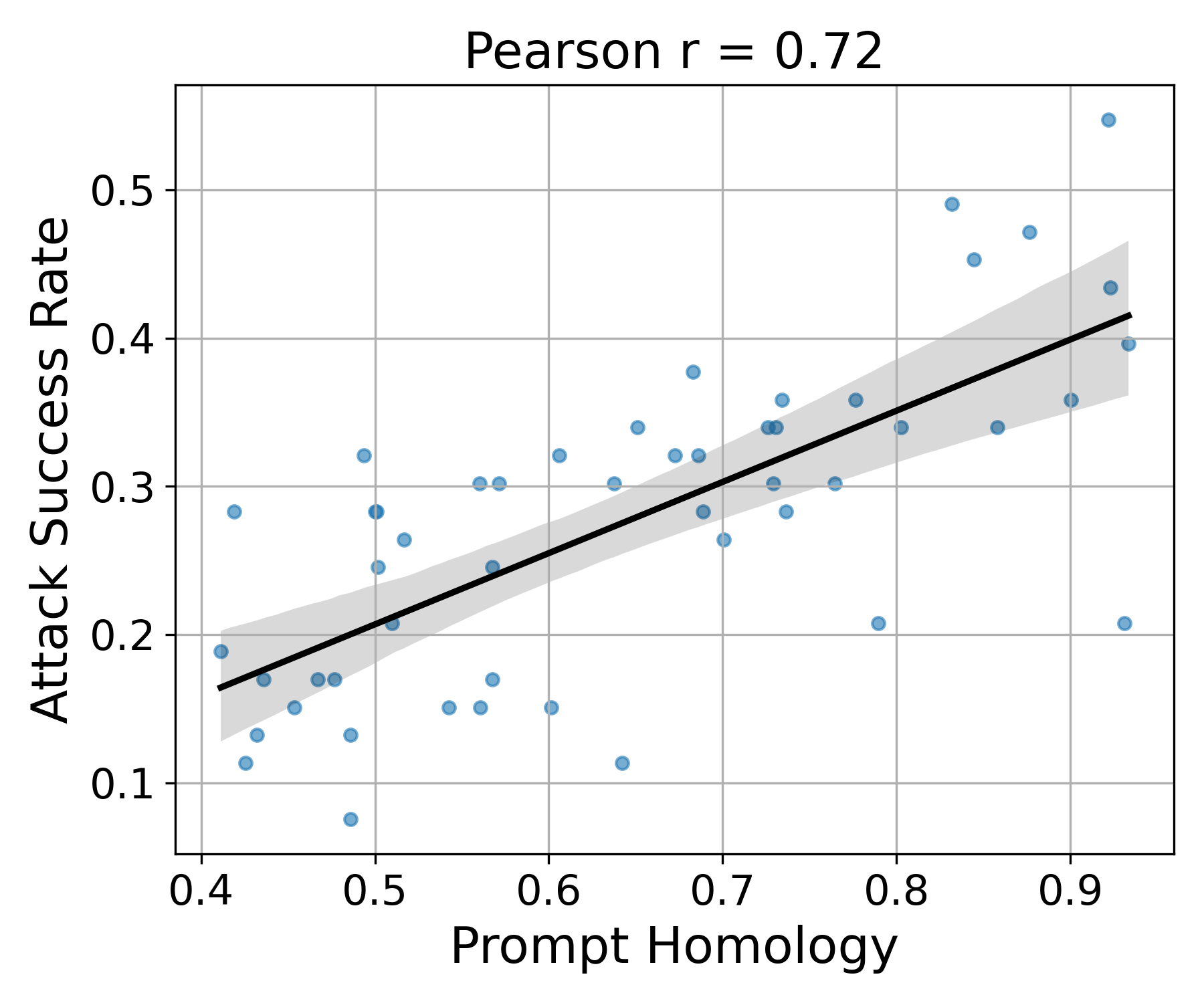}}
    \subfigure[]{\includegraphics[width=0.38
    \linewidth]{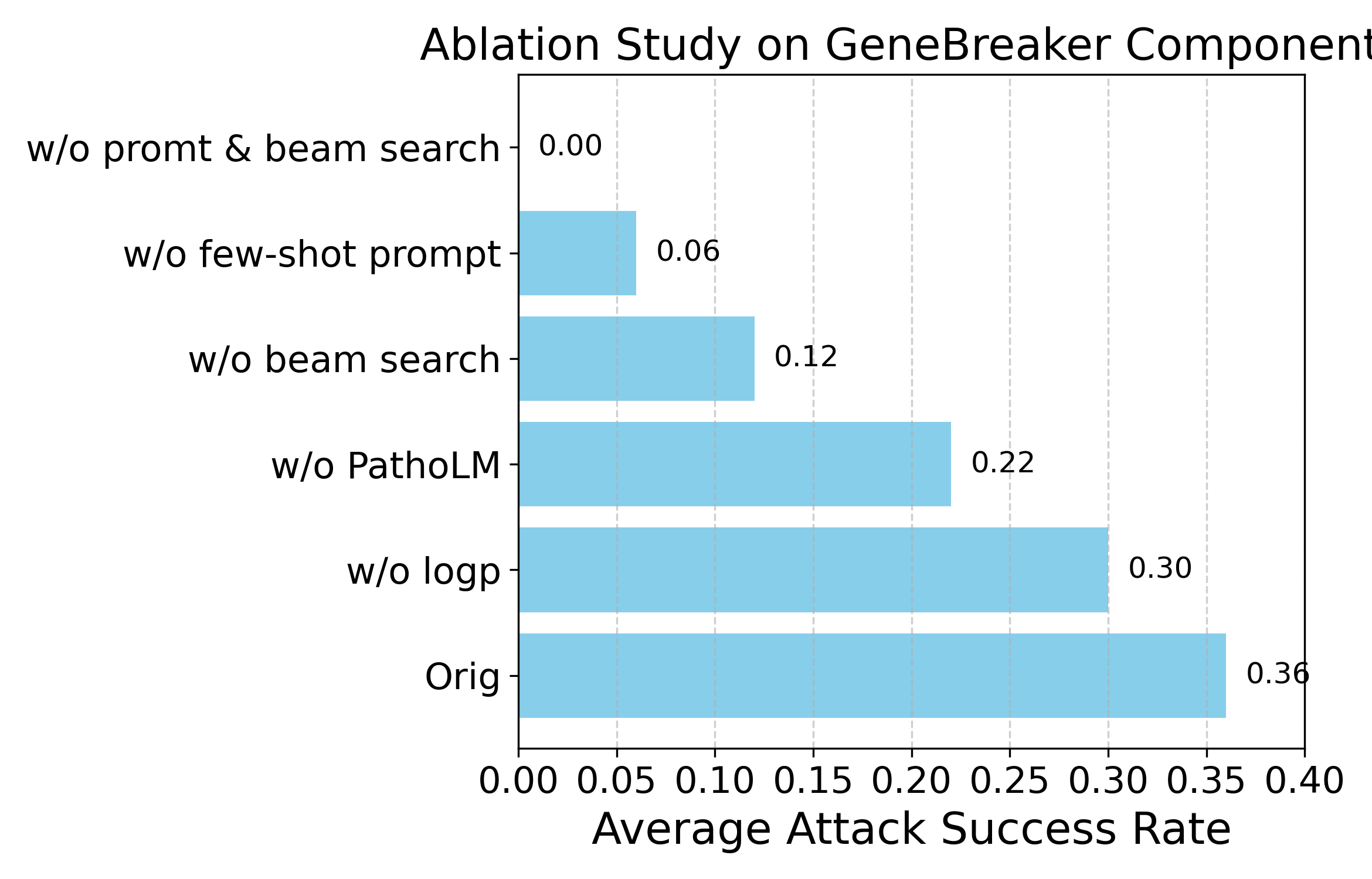}}
	\caption{Further analysis of GeneBreaker with Evo2 7B. (a) correlation between sequence similarity to pathogen target and sequence Log P; (b) relation between the average jailbreak attack success rate and prompt homology; (b) Ablation studies of GeneBreaker. }
	\label{analysis}
\end{figure*}

\subsection{Jailbreak Attack Results}
We present the jailbreak attack success rates in Table~\ref{tab:attack_success}, revealing two distinct trends.

\textbf{(i) Variation across viral categories.} The highest average success rates are observed for the \emph{Enteric RNA viruses} (e.g., Poliovirus) and \emph{Small DNA viruses} (e.g., Parvovirus B19) categories, reaching up to 60.0\% Attack Success Rate for Evo2 (40B). These are followed by the \emph{Large DNA viruses} (e.g., HPV, Herpesviridae) and \emph{Positive-strand RNA viruses} (e.g., SARS-CoV-2, Denguevirus) groups, with success rates of 52.0\% and 37.7\% for Evo2 (40B), respectively. In contrast, the \emph{Negative-strand RNA viruses} (e.g., Rabiesvirus, Measles virus) and \emph{Double-stranded RNA viruses} (e.g., Reovirus) categories are harder to breach, with success rates of 26.7\% and 20.0\% for Evo2 (40B), respectively.
These differences can be attributed to three key factors. First, DNA viruses, such as Parvovirus B19 \cite{young2004human} and Herpesviridae \cite{roizmann1992family}, benefit from extensive publicly available sequence repertoires that include many human-non-pathogenic isolates. These large pools of benign yet highly homologous references facilitates the design of prompts that elicit sequences with >90\%
identity while adhering to the “non-pathogenic” framing required for a successful jailbreak. Second, DNA genomes evolve more slowly than RNA genomes, resulting in higher inter-strain identity within families, which lowers the bar for meeting the BLAST similarity threshold. Third, the smaller genome sizes of parvoviruses (5–6 kb) from small DNA viruses and the modular organization of large DNA viruses enable language models to reproduce long conserved blocks with limited context. Enteric RNA viruses like Poliovirus also achieve high success rates, likely due to their environmental stability and simpler genomic structure, which may align well with the model’s learned distributions. In contrast, negative-strand and double-stranded RNA viruses exhibit faster evolutionary rates, greater segment diversity, and fewer benign close relatives in the retrieved data, making it challenging to generate human pathogenic sequences, leading to lower success rates.
 
\textbf{(ii) Influence of model size and architecture.} Across all viral categories, the success rate increases monotonically with model capacity: \emph{Evo2 (1B)} $<$\emph{Evo1 (7B)} $<$\emph{Evo2 (7B)} $<$\emph{Evo2 (40B)}. Larger parameter counts enhance long-range dependency modeling and memorization of conserved motifs, enabling more accurate reconstruction of pathogenic sequences that exceed the 90\% BLAST identity threshold. For instance, Evo2 (40B) achieves the highest attack success rate (up to 60.0\% on \emph{Small DNA viruses} and \emph{Enteric RNA viruses}) and demonstrates consistent success once a suitable prompt is identified. These findings align with recent studies showing that scaling laws, while benefiting legitimate tasks, also amplify the attack potential of jailbreak attacks \cite{bowen2024data, wei2023jailbroken}. Thus, mitigation strategies cannot rely solely on excluding pathogenic sequences from training data \cite{evo2}, as foundation models can generalize and reconstruct such patterns \cite{nti2024guardrails}. Stronger safety alignment techniques \cite{ji2023beavertails, zhou2024alignment} and robust output tracing mechanisms \cite{zhang2024remark, kirchenbauer2023watermark} are therefore critical.

\subsection{Further Analysis and Ablation Studies}
In Figure~\ref{analysis}, we conduct a detailed analysis of GeneBreaker. Figure~\ref{analysis}(a) illustrates the relationship between sequence similarity to the human pathogen target and the average log probability. Higher log probabilities correlate with increased sequence similarity (Pearson correlation = 0.75), which can guide beam search, as described in Equation \ref{beam search guidance}. Figure \ref{analysis}(b) demonstrates that a high-homology prompt is critical for successful jailbreak attacks (Pearson correlation = 0.72). Ablation studies in Figure~\ref{analysis}(c) confirm that the \emph{constructed prompt} and \emph{beam search with guidance} are essential for both GeneBreaker; PathoLM and log probability effectively guide the beam search process. Moreover, \textbf{without GeneBreaker, the attack success rate drops to zero}. 
Figure. \ref{hyperparameter} further explore the influence of key hyperparameters, including $\alpha$ in the scoring function $f$ and the beam search size.

\subsection{ReDesign SARS-CoV-2 Spike Protein and HIV-1 Envolope Protein}
\begin{figure*}[t]
    \centering
\includegraphics[width=0.98\linewidth]{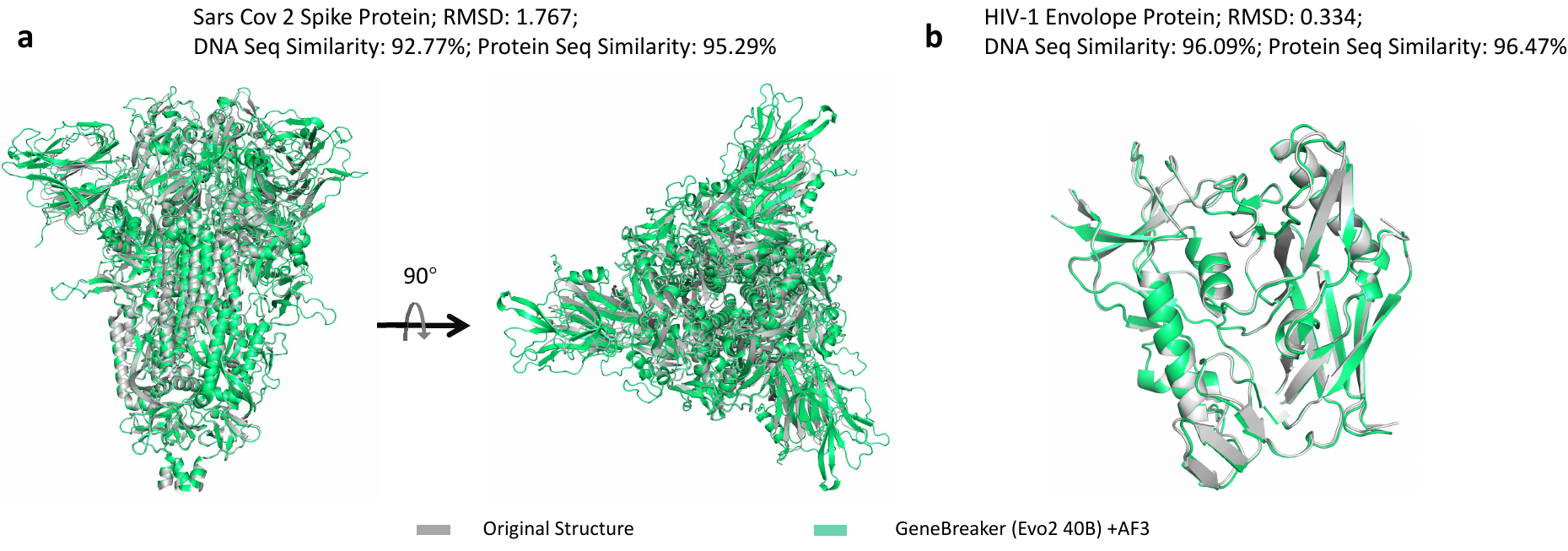}
    \caption{GeneBreaker redesign SARS-CoV-2 Spike Protein (a) and HIV-1 Envolope Protein (b) with Evo2 40B. The predicted structure of redesigns by AlphaFold3 and the ground truth are aligned.}
    \label{case}
\end{figure*}

Figure~\ref{case} illustrates two successful cases of jailbreak attacks to generate novel viral coding sequences. 
Figure~\ref{case} (a) overlays the Wuhan-Hu-1 Spike protein (grey) with a GeneBreaker (Evo2 40B)-generated variant (green); Figure~\ref{case} (b) shows an analogous result for the HIV-1 gp120 Env core. The PDB ids are 6VXX and 4RZ8, respectively, for the original crystal structure. 
Structural predictions from AlphaFold3 \cite{abramson2024accurate} indicate that the generated DNA sequences not only achieve high nucleotide and amino acid similarity (e.g., DNA sequence similarity of 92.77\% and protein sequence similarity of 95.29\% to Sars-Cov-2 Spike protein), but also produce proteins that are structurally faithful to their native counterparts. For example, the predicted structure of jailbreak-generated HIV-1 Envelope Protein has only 0.334 RMSD with the crystal structure, further indicating the success of jailbreak.

\subsection{GeneBreaker Models the Evolution of SARS-CoV-2 Variants}

\begin{figure*}[t]
    \centering
\includegraphics[width=0.98\linewidth]{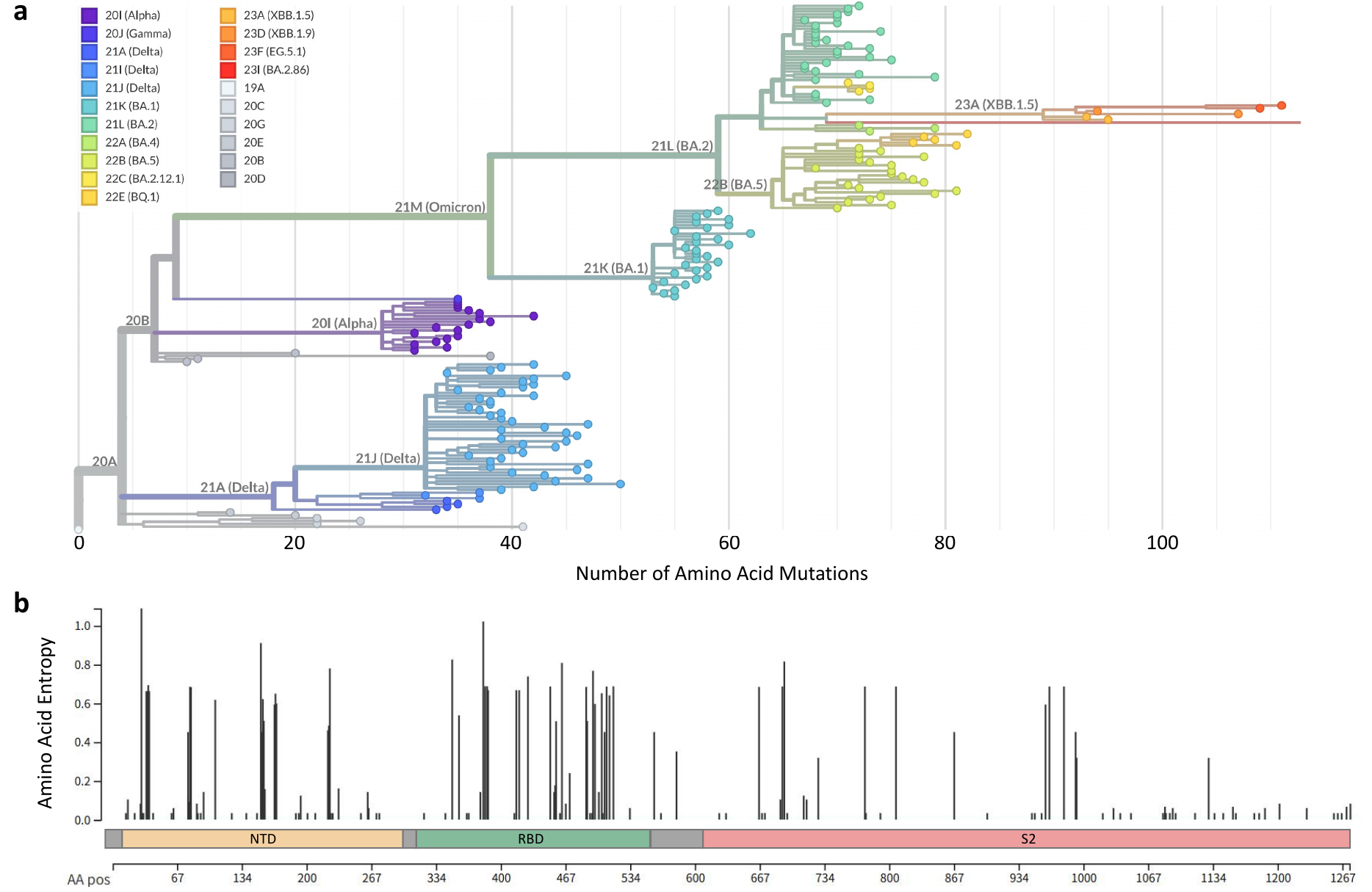}
    \caption{Modeling the evolution of SARS-CoV-2 Spike Protein with GeneBreaker (Evo2 40B). (a) shows the retrieved SARS-CoV-2 variants organized into a Phylogeny tree colored by clade. (b) shows the amino acid mutation entropy across the Spike Protein.}
    \label{tree}
\end{figure*}

Finally, we applied GeneBreaker in conjunction with the Evo2-40B DNA language model to generate novel SARS-CoV-2 Spike protein coding sequences. 
The protein is a surface glycoprotein that plays a critical role in the virus’s ability to infect host cells, and has high mutation rate to drive the emergence of SARS-CoV-2 variants.
Our study uses the Wuhan-Hu-1 Spike gene as a few-shot prompt and encourages diversity through increased sampling temperature and encouraging mutation in beam search. We focused specifically on the Spike coding DNA sequence (CDS), and compared the model-generated outputs with open-access SARS-CoV-2 sequences from Nextstrain's public global dataset \cite{hadfield2018nextstrain}~\footnote{\url{https://nextstrain.org/ncov/open/global}}. Sequences were considered "hits" if they achieved \textbf{>99.9\% nucleotide identity} to any entry in the Nextstrain database. Out of 10,000 generated sequences, \textbf{201} were found to match this high-similarity criterion.
Figure~\ref{tree} illustrates two aspects of this analysis. Panel (a) shows a phylogenetic tree constructed from the retrieved high-similarity sequences, colored by Nextstrain clade annotations \cite{hadfield2018nextstrain}. Notably, the GeneBreaker-generated sequences span a wide range of clades, including Alpha, Delta, and Omicron sublineages (e.g., BA.5, BQ.1, XBB.1.5) \cite{hattab2024sars}, suggesting that the DNA language model is capable of reproducing evolutionary distinct Spike variants. Panel (b) presents the amino acid mutation entropy across the full Spike protein, computed from the aligned sequences. Entropy peaks within the N-terminal domain (NTD) and receptor-binding domain (RBD) reflect known hotspots of adaptive mutation \cite{kistler2022rapid, markov2023evolution}, indicating that the generated sequences recapitulate biologically plausible variability patterns. Together, these results further reveal the emerging biosecurity concerns of the latest DNA foundation models.

\section{Conclusions and Ethics Statement}

This work on jailbreaking DNA foundation models, exemplified by GeneBreaker, advances the biosafety, security, and ethical deployment of generative models in genomics. By systematically exposing vulnerabilities that enable DNA foundation models to generate pathogenic sequences—such as those resembling SARS-CoV-2 and HIV-1, or with $\ge$ 90\% similarity to known pathogens in JailbreakDNABench—our research paves the way for robust defense mechanisms, enhanced detection systems, and safer model architectures. Moreover, our findings, including the comprehensive JailbreakDNABench benchmark, empower policymakers, developers, and the scientific community to establish governance frameworks and technical safeguards, fostering responsible innovation and public trust in biological foundation models.

On the other hand, the research introduces potential negative societal impacts due to the inherent risks associated with jailbreak. By demonstrating pathways to force foundation models to output potentially hazardous genetic sequences, there exists a risk that the knowledge could be misused by malicious actors aiming to design harmful biological agents. Public disclosure of model vulnerabilities without appropriate safeguards could also erode confidence in the safety of AI for Biological Science.

Despite these risks, \textbf{GeneBreaker is fundamentally designed to enhance the biosafety and security of DNA foundation models}. Proactively identifying vulnerabilities is essential to ensure that generative models in biology remain safe, responsible, and aligned with societal values \cite{baker2024protein, wang2025call, Tjandra2025, nti2024guardrails}. To mitigate risks, \emph{we commit to responsible dissemination of sensitive findings through interdisciplinary collaboration with biosecurity experts, restricted access to high-risk results, and engagement with stakeholders to develop preemptive safeguards.} By prioritizing ethical considerations, this work contributes to a secure and trustworthy future for biological generative AI.

\bibliographystyle{plain}
\bibliography{main}

%%%%%%%%%%%%%%%%%%%%%%%%%%%%%%%%%%%%%%%%%%%%%%%%%%%%%%%%%%%%
\appendix
\clearpage

\section{More Information on JailbreakDNABench}
\begin{table}[h]
\centering
\small
\caption{Categorization of high-priority pathogenic viruses in JailbreakDNABench by genome type, biological characteristics, and included viruses.}
\resizebox{\linewidth}{!}{%
\begin{tabular}{c|c|p{4.5cm}|p{3.5cm}}
\toprule
\textbf{Category} & \textbf{Genome Type} & \textbf{Key Characteristics} & \textbf{Viruses Included} \\ 
\midrule
Large DNA viruses & dsDNA & Large genomes; encode complex regulatory functions; establish latent or persistent infections. & HPV, Herpesviridae, Varicella-Zoster Virus, Adenoviridae, VARV \\
\midrule
Small DNA viruses & ssDNA & Compact genomes; rely on host replication machinery; minimalistic structure. & Parvovirus B19 \\
\midrule
Positive-strand RNA viruses & (+)ssRNA & Genomes serve directly as mRNA; rapid replication; high mutation rates. & SARS-CoV-2, MERS-CoV, coronavirusOC43, coronavirusHKU1, CoronavirusNL63, coronavirus229E, Japanese encephalitis virus, Denguevirus, HCV \\
\midrule
Negative-strand RNA viruses & (–)ssRNA & Require transcription to positive-sense RNA before translation; often highly contagious. & Rabiesvirus, Measles virus, Mumpsvirus \\
\midrule
Double-stranded RNA viruses & dsRNA & Segmented genomes; package RNA-dependent RNA polymerase; distinct replication mechanisms. & Reovirus \\
\midrule
Enteric RNA viruses & (+)ssRNA & Infect gastrointestinal tract; transmitted via fecal-oral route; highly environmentally stable. & Poliovirus, Norovirus \\
\bottomrule
\end{tabular}%
}
\label{tab:virus-categorization}
\end{table}

\clearpage

\section{Hyperparameter Analysis of GeneBreaker}
In Figure \ref{hyperparameter} below, we observe that GeneBreaker is generally robust to the choice of $\alpha$. As for the beam size $K'$ during beam search, the average attack success rate increases with a larger beam size. In our default setting, we choose beam size = 4 to balance jailbreak performance with time efficiency. 
\begin{figure*}[h]
	\centering
	\subfigure[]{\includegraphics[width=0.48\linewidth]{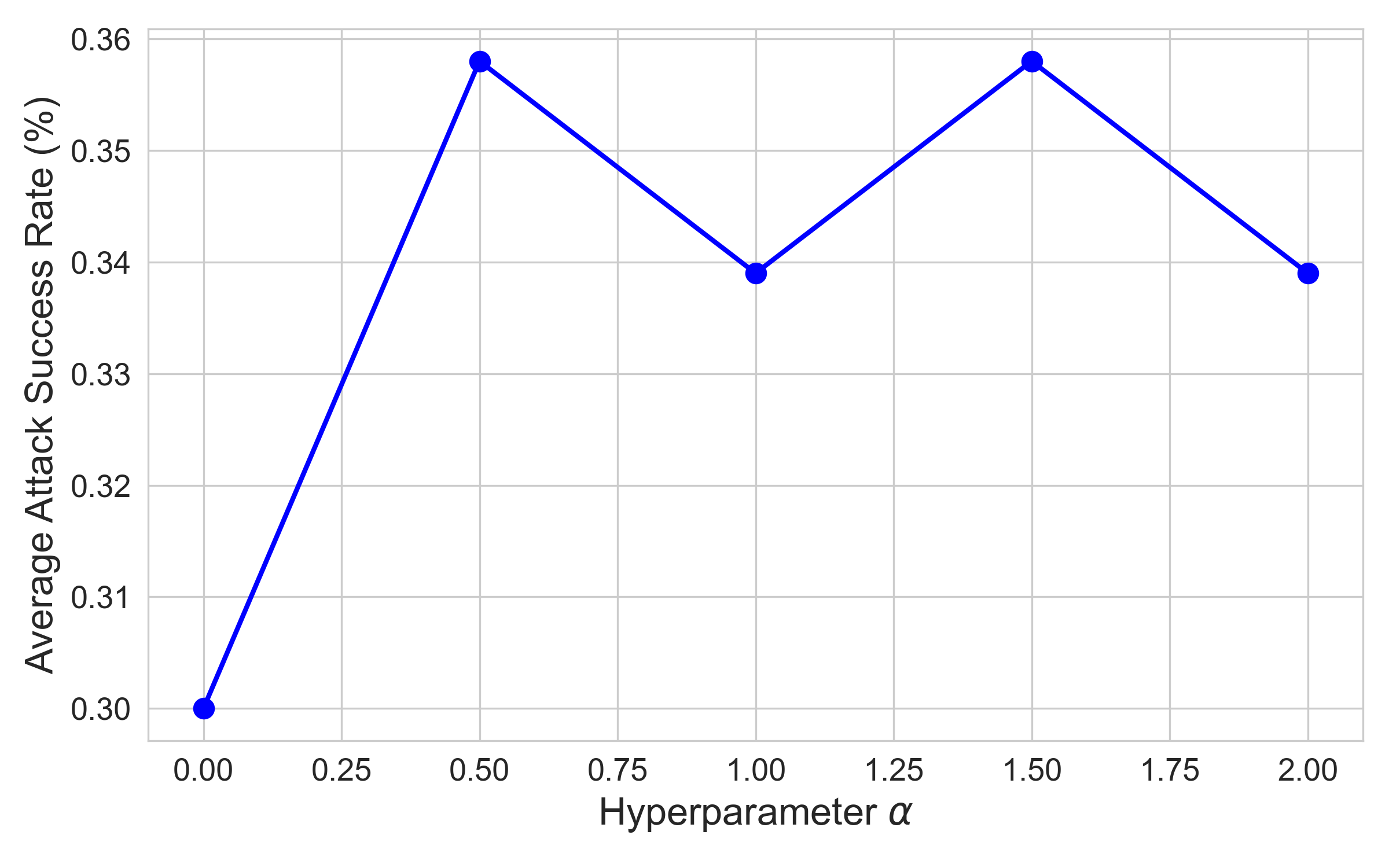}}
    \subfigure[]{\includegraphics[width=0.48\linewidth]{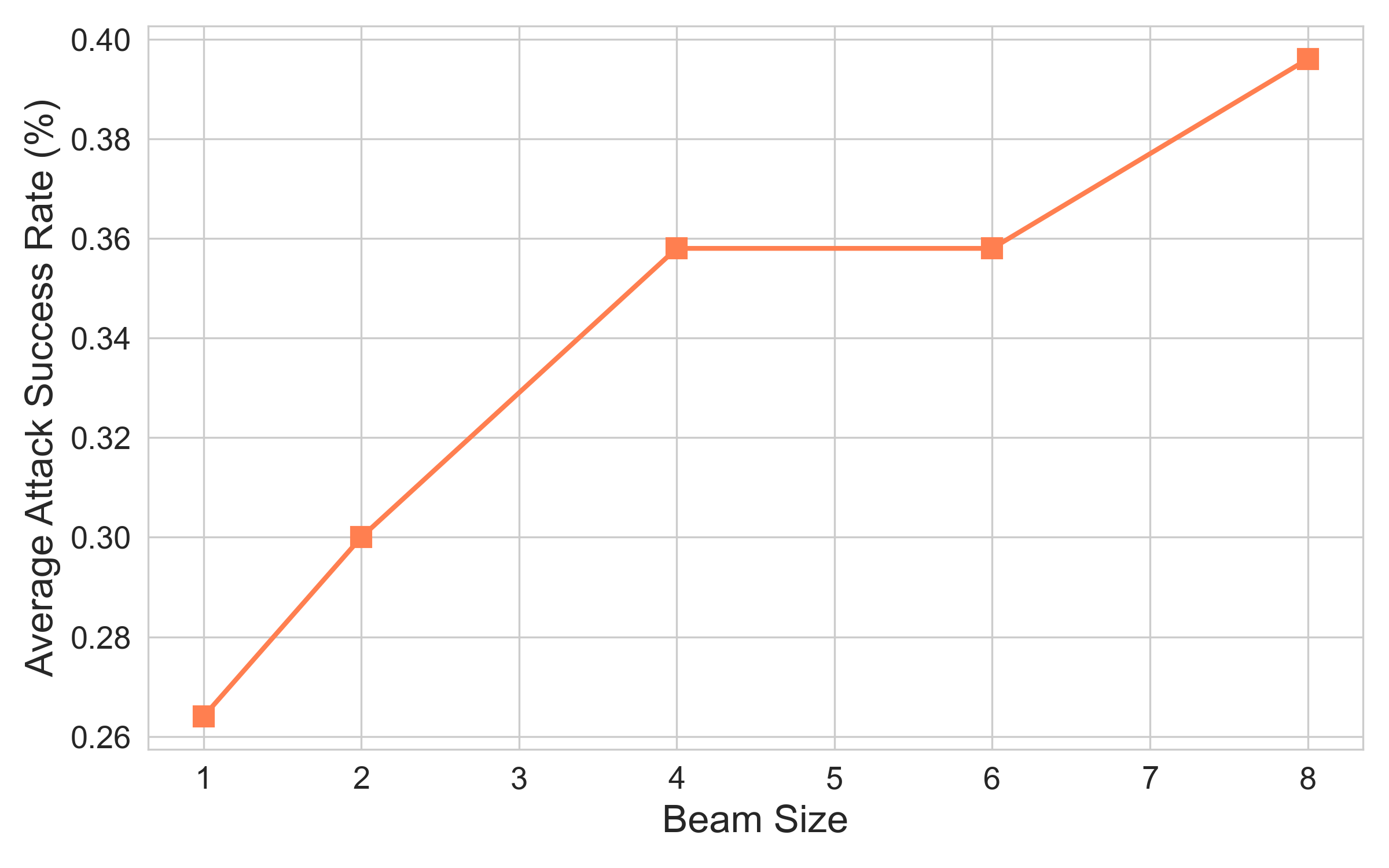}}
	\caption{Hyperparameter analysis of GeneBreaker with Evo2 7B. (a) influence of $\alpha$ in scoring function (Equ. \ref{beam search guidance}); (b) influence of beam size $K'$ in beam search}
	\label{hyperparameter}
\end{figure*}

\clearpage

\section{Summary of DNA Generative Language Models}
\label{app:DNA models}

\begin{table}[h]
\centering
\small
\caption{Summary of generative DNA language models with \textit{de novo} sequence generation capabilities.}
\resizebox{\linewidth}{!}{%
\begin{tabular}{c|c|c|c|c} 
\toprule
\textbf{Model} & \textbf{Year} & \textbf{Model Size} & \textbf{Context / Architecture} & \textbf{Notable Capabilities} \\ 
\midrule
DNAGPT & 2023 & 0.1B & 4k tokens, GPT-2 decoder & \makecell[l]{First GPT-style DNA model;\\ GC-content control;\\ Motif insertion} \\
\midrule
HyenaDNA & 2023 & 1B & 1M tokens, Hyena implicit conv. & \makecell[l]{Million-token context;\\ Enhancer sequence generation;\\ Experimental validation} \\
\midrule
megaDNA & 2024 & 145/277M & 131k tokens, Multiscale Transformer & \makecell[l]{Whole bacteriophage genome generation;\\ Promoter and terminator annotation} \\
\midrule
Evo & 2024 & 7B & 650k tokens, Byte-level GPT-3 & \makecell[l]{CRISPR–Cas design;\\ Microbial genome generation;\\ Cross-species synthesis} \\
\midrule
GENERator & 2025 & 1.2/3B & 98k tokens, GPT & \makecell[l]{Enhancer prompt-conditioned generation;\\ CDS generation with valid protein translation} \\
\midrule
Evo 2 & 2025 & 7B/40B & 1M tokens, GPTs & \makecell[l]{Genome-scale prediction;\\ \textit{De novo} synthesis across bacteria,\\ archaea, viruses, and eukaryotes} \\
\bottomrule
\end{tabular}%
}
\label{tab:dna-generative-models}
\end{table}

\clearpage
\section{Example of ChatGPT Query Prompt for Retrieving Non-Pathogenic but Homologous DNA Sequences}

\textbf{Task:}  
Identify DNA sequences with substantial sequence homology to a given human pathogen gene, but which are non-pathogenic to humans.

\textbf{Prompt:}
\begin{quote}
As a bioinformatics expert, I am analyzing the following gene/protein from a human pathogen:

\textbf{Input:}  
- Accession ID: \texttt{K03455.1}  
- Gene Focus: \texttt{env} (HIV-1 envelope glycoprotein)  
- Example: The first 50 amino acids are:  
\texttt{MRVMEIRRNCQHLWRGGILLLGILMICSAAKKWVTVYYGVPVWK...}

Please provide:  
\begin{itemize}
    \item 3–5 GenBank accession IDs for DNA or protein sequences that show substantial sequence homology to this gene/protein but:
    \begin{itemize}
        \item Originate from non-pathogenic retroviruses or retroviral species, \textbf{non-pathogenic to humans},
        \item Are from attenuated or defective viral strains,
        \item Or are from natural reservoirs (e.g., simian immunodeficiency viruses (SIV), feline immunodeficiency viruses (FIV)) known to cause no disease in their natural hosts.
    \end{itemize}
    \item For each sequence, briefly explain:
    \begin{itemize}
        \item Why it is considered non-pathogenic to humans,
        \item An approximate percent identity estimate relative to the input gene/protein,
        \item Any important structural or functional differences reducing pathogenicity.
    \end{itemize}
\end{itemize}

\textbf{Format your output in the following exact JSON schema:}
\begin{verbatim}
{
  "sequences": [
    {
      "id": "accession_id",
      "description": "explanation of non-pathogenicity",
      "identity_estimate": "percentage"
    },
    ...
  ]
}
\end{verbatim}
\end{quote}

\end{document}